# A tunable coupling scheme for implementing high-fidelity two-qubit gates

Fei Yan[1],* Philip Krantz[1], Youngkyu Sung[1], Morten Kjaergaard[1], Dan Campbell[1],
Joel I.J. Wang[1], Terry P. Orlando[1], Simon Gustavsson[1], and William D. Oliver[1,2,3]

[1]*Research Laboratory of Electronics, Massachusetts Institute of Technology, Cambridge, MA 02139, USA*
[2]*MIT Lincoln Laboratory, 244 Wood Street, Lexington, MA 02420, USA*
[3]*Department of Physics, Massachusetts Institute of Technology, Cambridge, MA 02139, USA*

The prospect of computational hardware with quantum advantage relies critically on the quality of quantum gate operations. Imperfect two-qubit gates is a major bottleneck for achieving scalable quantum information processors. Here, we propose a generalizable and extensible scheme for a two-qubit coupler switch that controls the qubit-qubit coupling by modulating the coupler frequency. Two-qubit gate operations can be implemented by operating the coupler in the dispersive regime, which is non-invasive to the qubit states. We investigate the performance of the scheme by simulating a universal two-qubit gate on a superconducting quantum circuit, and find that errors from known parasitic effects are strongly suppressed. The scheme is compatible with existing high-coherence hardware, thereby promising a higher gate fidelity with current technologies.

Recent developments of quantum information processor architectures have been focusing on scalability [1–5]. High-quality gate operations is one of the key performance indicators for these intermediate-scale quantum processors [6]. Since the gate performance ultimately determines if a device can exhibit quantum advantage, the development of high-quality gates in these systems becomes an imperative. Improving gate fidelity significantly reduces the overhead needed for implementing gate-based quantum error correcting codes and enhances the performance of quantum simulations. The major limiting factor for quantum gate operations today is the relatively faulty two-qubit gate. Therefore, improving two-qubit gate fidelity is of high priority to realize large-scale quantum processors.

In general, there are two sources of gate errors: decoherence (stochastic) and non-ideal interactions (deterministic). The latter includes parasitic coupling, leakage to non-computational states, and control crosstalk. As one example of parasitic coupling, the next-nearest-neighbor (N.N.N.) coupling is a phenomenon commonly seen in many systems, including Rydberg atoms [4, 7], trapped ions [5, 8], semiconductor spin qubits [9, 10], and superconducting qubits [11, 12]. Often, the N.N.N. coupling is considered spurious and introduces unwanted interactions between qubits that are meant to be unconnected.

At the same time, a coupling switch can help mitigate the problem of frequency crowding that exacerbates the effect from non-ideal interactions. Prototypes of a tunable coupler have been demonstrated extensively in superconducting quantum circuits [13–21]. However, these additional elements often add architectural complexity, as well as open a new channel for decoherence and crosstalk. Among them, the gmon design [19] is a successful example that exhibits a two-qubit gate fidelity limited predominantly by decoherence. However, the qubits' coherence times in gmon circuits are reduced by the tunable coupler in comparison with its predecessor, the xmon design [11].

In this work, we propose a simple and broadly applicable scheme for a tunable coupler and use it as a switch for implementing high-fidelity two-qubit gates. The approach is based on a generic three-body system with exchange-type interaction. A central component, the coupler, frequency tunes the virtual exchange interaction between two qubits and features a critical bias point, at which the exchange interaction offsets the direct qubit-qubit (N.N.N.) coupling, effectively turning off the net coupling. Two-qubit gate operations are executed by operating the coupler in the dispersive regime, strongly suppressing leakage to the coupler's excited states. We simulate the iSWAP gate based on an existing high-coherence superconducting quantum hardware in our group [22] and elsewhere [11]. We find that gate errors due to parasitic effects diminish drastically with increased gate time (decreased interaction amplitude). A gate fidelity above 99.999% can be achieved in 100 ns in the absence of decoherence. The utilization of N.N.N. coupling, the compatibility with high-coherence architecture, and the strong suppression of parasitic effects all make our scheme a viable choice for the long term as coherence times continue to improve.

We consider a generic system consisting of a chain of three modes with exchange coupling between nearest and next-nearest neighbors, as outlined in Fig. 1(a). The two qubits ($\omega_1$ and $\omega_2$) each couple to a center tunable coupler ($\omega_c$) with a coupling strength $g_j$ ($j = 1, 2$), as well as to each other with a coupling strength $g_{12}$. The nearest-neighbor (N.N.) coupling is generally stronger than the N.N.N. coupling, $g_j > g_{12} > 0$. Without loss of generality, we begin our analysis with a two-level Hamiltonian,

$$H = \sum_{j=1,2} \frac{1}{2}\omega_j \sigma_j^z + \frac{1}{2}\omega_c \sigma_c^z + \sum_{j=1,2} g_j \left(\sigma_j^+ \sigma_c^- + \sigma_j^- \sigma_c^+\right)$$
$$+ g_{12}\left(\sigma_1^+ \sigma_2^- + \sigma_2^- \sigma_1^+\right), \qquad (1)$$

where $\sigma_\lambda^z$, $\sigma_\lambda^+$ and $\sigma_\lambda^-$ ($\lambda = 1, 2, \mathrm{c}$) are, respectively,

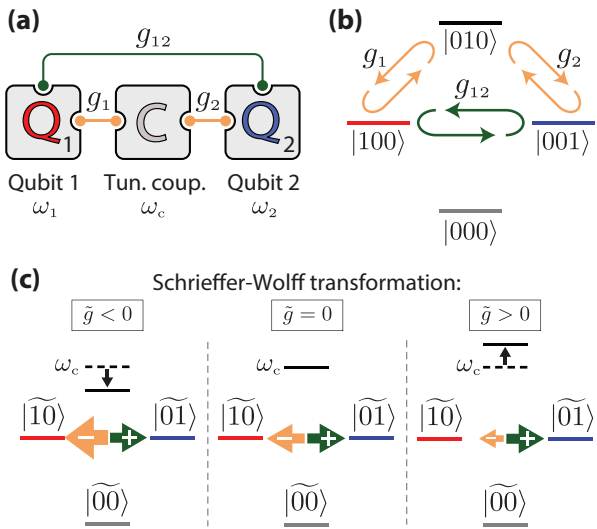

FIG. 1. **(a)** Sketch of a generic three-body system in a chain geometry, where the center mode is a tunable coupler. **(b)** Level diagram of the ground and one-excitation states of the system. The ket symbol follows the chain order $|\omega_1, \omega_c, \omega_2\rangle$. The round-trip arrows indicate N.N. (orange) and N.N.N. (green) coupling. **(c)** Level diagrams of the reduced two-qubit system after Schrieffer-Wolff transformation of the level diagram in (b). Each figure corresponds to the case of an effective negative (left), zero (center), and positive (right) net coupling, $\tilde{g}$. The double-headed arrows indicate the sign and magnitude of the coupling. In this example, the N.N.N. coupling (green) is positive and fixed. The N.N. coupling (orange) is negative and tunable with the coupler energy (solid black line).

the Pauli-$Z$, raising and lowering operators defined in the eigenbasis of the corresponding mode. We assume that both qubits are negatively detuned from the coupler, $\Delta_j \equiv \omega_j - \omega_c < 0$, and that the coupling is dispersive, $g_j \ll |\Delta_j|$ $(j = 1, 2)$. Fig. 1(b) sketches the level structure of this system. The two qubits interact through two channels, the direct N.N.N. coupling and the indirect coupling via the coupler. The latter is sometimes called virtual exchange interaction [23], which can be approximated by the Schrieffer-Wolff transformation (SWT) $U = \exp\left[\sum_{j=1,2} \frac{g_j}{\Delta_j}(\sigma_j^+ \sigma_c^- - \sigma_j^- \sigma_c^+)\right]$ [24]. The transformation decouples the coupler from the system up to second order in $\frac{g_j}{\Delta_j}$, resulting in an effective two-qubit Hamiltonian,

$$\widetilde{H} = \sum_{1,2} \frac{1}{2}\widetilde{\omega}_j \sigma_j^z + \left[\frac{g_1 g_2}{\Delta} + g_{12}\right](\sigma_1^+ \sigma_2^- + \sigma_2^- \sigma_1^+), \quad (2)$$

where $\widetilde{\omega}_j = \omega_j + \frac{g_j^2}{\Delta_j}$ is the Lamb-shifted qubit frequency and $1/\Delta = (1/\Delta_1 + 1/\Delta_2)/2 < 0$. Here, we have also assumed that the coupler mode remains in its ground state at all times.

The combined term inside the square brackets in Eq. (2) represents the total effective qubit-qubit coupling $\tilde{g}$. It can be adjusted by the coupler frequency through $\Delta$, as well as $g_1$ and $g_2$, both of which may be implicitly dependent on $\omega_c$. Thus, $\tilde{g}$ is a function of $\omega_c$ in general. Moreover, since $1/\Delta < 0$, the first term in the square brackets – the virtual exchange interaction – is negative. This enables a competition between the positive direct coupling and the negative indirect coupling. As illustrated in Fig. 1(c), $\tilde{g}(\omega_c)$ can be tuned negative when the coupler frequency is decreased, or positive when the coupler frequency is increased. Most importantly, since the tunability is continuous, one can always find a critical value $\omega_c^{\text{off}}$ at which the two terms cancel out and thereby turn off the coupling, i.e., $\tilde{g}(\omega_c^{\text{off}}) = 0$, as long as permitted by the bandwidth of the coupler. Note that the dispersive-limit condition is only an ideal requirement. In systems with considerably greater $g_{12}$, it is still possible to find such an $\omega_c^{\text{off}}$ in the weakly dispersive regime $(g_j < |\Delta_j|)$.

The tunable coupler is used as a switch by biasing its frequency at $\omega_c^{\text{off}}$ during idling periods. To activate the two-qubit interaction, one tunes the coupler frequency to a desired value $\omega_c^{\text{on}}$, yielding a finite $\tilde{g}(\omega_c^{\text{on}})$. The advantages of this scheme are three-fold: (i) The scheme solves the problem of unwanted N.N.N. coupling by incorporating it into the switch. Although N.N.N. coupling is not currently a major issue for quantum information processing architectures, it is an unavoidable consequence that all scalable modalities either have or will face. (ii) By operating the coupler in the dispersive limit, parasitic effects from higher-order terms that are ignored after SWT (Eq. (2)) are strongly suppressed, leading to higher two-qubit gate fidelity. (iii) In addition, a two-qubit gate can be performed by modulating only the coupler frequency while leaving the qubits unperturbed during the operation. For example, if the two qubits are resonant, an iSWAP gate can be implemented by turning on their coupling for a requisite amount of time. During this process, the control Hamiltonian $\sigma_c^z$ commutes with the qubits' degrees of freedom within the dispersive approximation, causing reduced leakage to the non-computational (coupler) state. The non-adiabatic effect in this case is suppressed by the relatively large qubit-coupler detuning $(\Delta_j)$, allowing a shorter gate time and therefore, reduced decoherence error.

The details of how to implement the scheme in a particular modality depends on the system parameters and controllability. For the remainder of the paper, we will focus on an implementation using superconducting qubits and numerically demonstrate the viability of our scheme. Consider a three-mode circuit $(\omega_1$-$\omega_c$-$\omega_2)$ outlined in Fig. 2(a), where each pair of modes are capacitively connected. Capacitive coupling with superconducting qubits is advantageous in preserving coherence times and com-

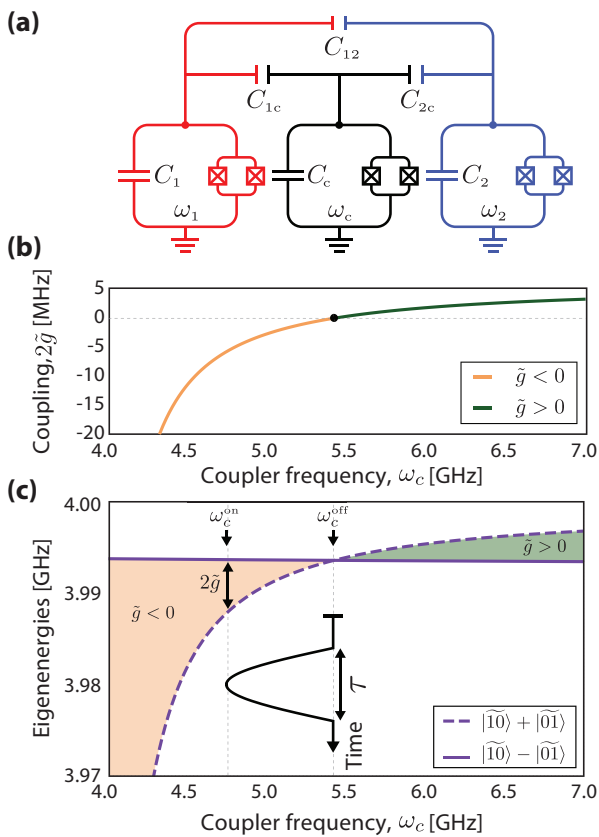

**(a)**

**(b)**

**(c)**

FIG. 2. **(a)** Circuit diagram of a superconducting circuit implementing a tunable coupler. Each mode is constructed by a tunable transmon qubit. **(b)** The $\omega_c$-dependence of $2\widetilde{g}$. The crossing at $2\widetilde{g}=0$ (black dot) indicates the switch-off bias $\omega_c^{\text{off}}$. **(c)** Calculated eigenenergies of the one-excitation manifold as a function of the coupler frequency. The parameters used are $C_1 = 70\,\text{fF}$, $C_2 = 72\,\text{fF}$, $C_c = 200\,\text{fF}$, $C_{1c} = 4\,\text{fF}$, $C_{2c} = 4.2\,\text{fF}$, $C_{12} = 0.1\,\text{fF}$, $\omega_1 = \omega_2 = 4\,\text{GHz}$. We intentionally create a 5-10% variation between $C_1$ and $C_2$ as well as between $C_{1c}$ and $C_{2c}$ to emulate fabrication variation (not a requirement for proof-of-concept). Since the two qubit modes are degenerate, the eigenstates are symmetric (solid line) and anti-symmetric (dashed line) combination of their wavefunctions, and the energy gap (shaded) corresponds to the effective coupling $2\widetilde{g}$ shown in (b). The inset illustrates the pulse that turns the switch on and off, executing an iSWAP gate.

patible with 3D integration [25]. In general, each mode represents a superconducting quantum nonlinear oscillator formed by a dominant capacitance ($C_1$, $C_2$, $C_c$) and a nonlinear inductance in parallel. $C_1$, $C_2$ and $C_c$ are of the same order of magnitude. Candidates for circuit implementation include single-junction or tunable transmons [26, 27], capacitively shunted flux qubits [28, 29], or capacitively shunted fluxonium qubits [30]. The center mode is used as a tunable coupler, which can be conveniently implemented with any flux-tunable circuit in which the resonance frequency can be tuned *in situ* by a time-dependent magnetic flux threading the coupler loop. Whether the qubit frequencies need to be tunable depends on the kind of gate scheme to be implemented. Fixed-frequency qubits can be equipped with the cross-resonance gate [31] or the parametrically driven gate [32]. In the example shown, we choose three tunable transmons qubits. We note that both the qubit-coupler capacitances $C_{jc}$ ($j = 1, 2$) and the qubit-qubit capacitance $C_{12}$ are small compared to any of $C_1$, $C_2$ and $C_c$, so the couplings are perturbative. Quantizing the circuit [33, 34], we obtain the system Hamiltonian in Eq. (1) with coupling terms

$$g_j \approx \frac{1}{2}\,\frac{C_{jc}}{\sqrt{C_j C_c}}\,\sqrt{\omega_1 \omega_c}\,, \qquad j = 1, 2 \qquad (3)$$

$$g_{12} \approx \frac{1}{2}\left[\frac{C_{12}}{\sqrt{C_1 C_2}} + \frac{C_{1c} C_{2c}}{\sqrt{C_1 C_2 C_c^2}}\right]\sqrt{\omega_1 \omega_2}\,. \qquad (4)$$

The qubit-qubit (N.N.N.) coupling $g_{12}$ has two contributions. The first term in the brackets in Eq. (4) is from the *direct* capacitive connection between the red and blue nodes in Fig. 2(a). The second term is from the *indirect* capacitive connection via the intermediate capacitance network formed by $C_{1c}$, $C_{2c}$ and $C_c$.

Since transmon qubits have weak anharmonicity, we generalize our model by including multilevels and counter-rotating terms [35]. Using Eqs. (3-4), we obtain the effective qubit-qubit coupling strength,

$$\widetilde{g} = \frac{1}{2}\left[\frac{\omega_c}{2\Delta}\eta - \frac{\omega_c}{2\Sigma}\eta + \eta + 1\right]\frac{C_{12}}{\sqrt{C_1 C_2}}\,\sqrt{\omega_1 \omega_2}\,, \qquad (5)$$

$\eta = C_{1c} C_{2c}/C_{12} C_c$, $1/\Sigma = (1/\Sigma_1 + 1/\Sigma_2)/2$ and $\Sigma_j = \omega_j + \omega_c$. The four terms represents respectively the coupling strength of (i) the virtual exchange interaction via the state $|010\rangle$ (indirect qubit-qubit coupling); (ii) the virtual exchange interaction via the state $|111\rangle$ (indirect qubit-qubit coupling); (iii) the capacitive coupling via the intermediate capacitance network (direct qubit-qubit coupling, indirect connection); (iv) the direct capacitive coupling between nodes (direct qubit-qubit coupling, direct capacitive connection). In practice, the N.N.N. capacitive connection is usually much weaker than the N.N. coupling ($C_{12} \ll C_{1c}, C_{2c}$). However, since the virtual interaction is a second-order effect, these four terms can have the same order of magnitude in their strength. For a realistic example, $\omega_1 = \omega_2 = 4\,\text{GHz}$, $\omega_c = 5\,\text{GHz}$, $\Delta = -1\,\text{GHz}$, $C_1 = C_2 = C_c = 100\,\text{fF}$, $C_{1c} = C_{2c} = 1\,\text{fF}$, $C_{12} = 0.02\,\text{fF}$ (similar device parameters as measured in Ref. [11]). The resulting dimensionless coupling strength from each contribution is (i) -1.25, (ii) -0.14, (iii) 0.5 and (iv) 1.0. We note that the role of last three terms is typically disregarded in treatments of circuits similar to that in Fig. 2(a) [23, 36]. However, as we show here, a careful inclusion of these three terms leads to important and non-negligible effects.

Since the qubits are negatively detuned from the coupler, we have $\frac{\omega_c}{2\Delta} - \frac{\omega_c}{2\Sigma} + 1 \leq 0$ (combined effect of (i), (ii) and (iii)), where the equality holds when the coupler frequency goes to infinity. Surprisingly, implementing the described circuit geometry with superconducting qubits inherently guarantees a solution for $\omega_c^{off}$ where the switch is off given any reasonable value of $C_{12}$. This makes our scheme widely applicable.

To quantify the performance, we numerically simulate an iSWAP gate. An iSWAP gate can be performed by executing half of an exchange period when $\widetilde{|01\rangle}$ and $\widetilde{|10\rangle}$ are degenerate (Fig. 1(c)). However, we emphasize that other types of two-qubit gates, such as a controlled-phase (C-phase) gate [37] or a parametrically driven gate [32], are also compatible with our scheme. We first calculate the values $\widetilde{g}(\omega_c)$ (Fig. 2(b)) by solving the system Hamiltonian (Fig. 2(c)). Hence, we identify the zero-coupling bias $\omega_c^{off}$. In experiments, one may calibrate $\omega_c^{off}$ by measuring vacuum Rabi oscillations while sweeping the coupler frequency. Next, we apply a cosine pulse of duration $\tau$ (Fig. 2(c) inset) that modulates $\omega_c$ and turns on the qubit-qubit interaction. For an iSWAP gate, the time-integral of the effective coupling satisfies $\int_0^\tau 2\widetilde{g}(t)\,dt = 1/2$. The final state is tomographically analyzed after correcting the dynamic phase of each qubit.

We performed the same protocol with various gate lengths and with options for including $T_1$ energy relaxation (uniform for all three modes) and quasistatic flux noise. First, the reduction of fidelity from quasistatic flux noise in the coupler loop is negligible ($< 10^{-6}$) assuming a typical flux fluctuation of $10\,\mu\Phi_0$ ($\Phi_0$ is the superconducting flux quantum), because when the switch is on, the sensitivity of the coupling $\delta\widetilde{g}/\delta\omega_c \approx \widetilde{g}/\Delta$ is reduced in the dispersive regime. Second, the gate infidelity (the error per gate) $\epsilon$ due to energy relaxation follows $\epsilon = \tau/T_1$ (We use 16 linearly independent input states when performing process tomography [38], and there is a prefactor difference from that estimated in randomized benchmarking [39]). Third, gate error due to effects other than decoherence drops quickly with increased gate time, because the major contributions to gate error are from higher-order parasitic couplings and have a stronger power-law dependence on the interaction strength $\widetilde{g}$. The last two contributions are illustrated in Fig. 3, where their crossings indicate the optimal operating point under our scheme given certain $T_1$ values. For example, if $T_1 = 10\,\mu s$, the optimal gate time and gate fidelity are $\tau^* = 35$ ns and $\epsilon^* = 3 \times 10^{-3}$ (total error: $6 \times 10^{-3}$), which is comparable to the state-of-the-art results based on tunable couplers [19]. However, the circuit model in our example is compatible with a simpler architecture, such as xmon qubits which have been demonstrated with reproducibly high $T_1$ values (20-40 $\mu s$) across the chip and low crosstalk [11]. Recent developments have also shown $T_1$ close to $100\,\mu s$ with a similar architecture [22]. Given $T_1 = 100\,\mu s$, the gate error at the optimal op-erating point (in this case 46 ns) is $5 \times 10^{-4}$ using our scheme. Future advances in materials and fabrication techniques will likely continue to enhance coherence. Assuming $T_1 = 1$ ms, our scheme can further lower the error rate to $6 \times 10^{-5}$ in 66 ns. The above analysis illustrates that our scheme can efficiently take advantage of improvements in coherence times with only small overhead in gate time.

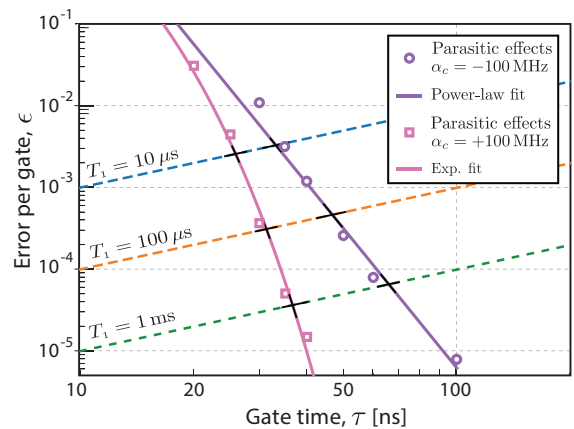

FIG. 3. Relation between error per gate $\epsilon$ and gate length $\tau$ due to energy relaxation and other parasitic effects respectively. Purple circles are simulation results in the absence of decoherence (bare dynamics) and using a negative coupler anharmonicity $\alpha_c = -100$ MHz, consistent with a transmon design. The solid purple line is a power-law fit, $\epsilon = 1.5 \times 10^6\,(\tau/1\text{ns})^{-5.7}$. Pink squares are simulation results in the absence of decoherence and using a positive coupler anharmonicity, $\alpha_c = +100$ MHz, consistent with a capacitively shunted flux qubit. The solid pink line is an exponential fit, $\epsilon = 73 \exp(-0.4\,\tau/1\text{ns})$. Dashed lines are calculated errors from $T_1$ process only (different $T_1$ values assumed). The crossings between the curves in the noise-free case and the $T_1$ case indicate approximately the break-even point for the optimal gate time and gate fidelity.

We further find that the remaining gate errors are mainly caused by a parasitic partial C-phase operation induced by high-order coupling between state $|020\rangle$ and $|101\rangle$. There are several approaches to eliminate this unwanted effect. One solution is to use a coupler mode with slightly positive anharmonicity to separate the two levels further apart. Here, anharmonicity is defined as the frequency difference between 1-2 and 0-1 transitions, i.e., $\alpha = \omega_{12} - \omega_{01}$. A potential candidate implementing such a design is the capacitively shunted flux qubit [29]. Simulation results with the same configuration but a coupler anharmonicity $\alpha_c = +100$ MHz show significant improvement compared to the case of $\alpha_c = -100$ MHz (Fig. 3). The remaining errors are largely due to leakage to the excited states of the coupler, evident as the exponential dependence on the gate length. An alternative solution is to perform the gate in the positive-$\tilde{g}$ regime. By turning up $\omega_c$ and entering deeper into the dispersive regime, unwanted effects can be suppressed. However, the gate speed is limited by the direct coupling ($C_{12}$). In the future, engineering a reproducible and stronger N.N.N. coupling can further empower this scheme. Finally, using optimized pulse shaping techniques [40] with our scheme can mitigate gate error from leakage and further improve fidelity.

In conclusion, we propose a simple and generic scheme for a coupler switch. The coupler can laisbe turned off completely by offsetting the direct qubit-qubit coupling with the virtual exchange interaction via the coupler. By operating the coupler in the dispersive regime, gate errors arising from non-ideal dynamics can be effectively suppressed. We demonstrate these properties by numerically simulating the scheme in a superconducting circuit. Our results suggest the performance of our scheme is mainly limited by $T_1$. Therefore, our scheme is viable in the long term as coherence times continue to improve.

We thank Andreas Bengtsson, Andrew Kerman, Zhirong Lin, Danna Rosenberg, Gabriel Samach, Mollie Kimchi-Schwartz, Michelle Wang for insightful discussions, and Mirabella Pulido for generous assistance. This research was funded in part by the Assistant Secretary of Defense for Research & Engineering via MIT Lincoln Laboratory under Air Force Contract No. FA8721-05-C-0002; by the U.S. Army Research Office Grant No. W911NF-14-1-0682; and by the National Science Foundation Grant No. PHY-1720311. Y.S. acknowledges support from the Korea Foundation for Advanced Studies (KFAS). M.K. acknowledges support from the Carlsberg Foundation. The views and conclusions contained herein are those of the authors and should not be interpreted as necessarily representing the official policies or endorsements of the US Government.

---

* fyan@mit.edu

[1] J. Kelly, R. Barends, A. G. Fowler, A. Megrant, E. Jeffrey, T. C. White, D. Sank, J. Y. Mutus, B. Campbell, Y. Chen, Z. Chen, B. Chiaro, A. Dunsworth, I. C. Hoi, C. Neill, P. J. O'Malley, C. Quintana, P. Roushan, A. Vainsencher, J. Wenner, A. N. Cleland, and J. M. Martinis, Nature **519**, 66 (2015).

[2] C. Song, K. Xu, W. Liu, C.-p. Yang, S.-B. Zheng, H. Deng, Q. Xie, K. Huang, Q. Guo, L. Zhang, P. Zhang, D. Xu, D. Zheng, X. Zhu, H. Wang, Y.-A. Chen, C.-Y. Lu, S. Han, and J.-W. Pan, Phys. Rev. Lett. **119**, 180511 (2017).

[3] J. S. Otterbach, R. Manenti, N. Alidoust, A. Bestwick, M. Block, B. Bloom, S. Caldwell, N. Didier, E. S. Fried, S. Hong, P. Karalekas, C. B. Osborn, A. Papageorge, E. C. Peterson, G. Prawiroatmodjo, N. Rubin, C. A. Ryan, D. Scarabelli, M. Scheer, E. A. Sete, P. Sivarajah, R. S. Smith, A. Staley, N. Tezak, W. J. Zeng, A. Hud-

son, B. R. Johnson, M. Reagor, M. P. da Silva, and C. Rigetti, arXiv preprint arXiv:1712.05771 (2017).

[4] H. Bernien, S. Schwartz, A. Keesling, H. Levine, A. Omran, H. Pichler, S. Choi, A. S. Zibrov, M. Endres, M. Greiner, V. Vuletic, and M. D. Lukin, Nature **551**, 579 (2017).

[5] J. Zhang, G. Pagano, P. W. Hess, A. Kyprianidis, P. Becker, H. Kaplan, A. V. Gorshkov, Z. X. Gong, and C. Monroe, Nature **551**, 601 (2017).

[6] J. Preskill, arXiv preprint arXiv:1801.00862 (2018).

[7] M. Viteau, M. G. Bason, J. Radogostowicz, N. Malossi, D. Ciampini, O. Morsch, and E. Arimondo, Phys. Rev. Lett. **107**, 060402 (2011).

[8] J. W. Britton, B. C. Sawyer, A. C. Keith, C.-C. J. Wang, J. K. Freericks, H. Uys, M. J. Biercuk, and J. J. Bollinger, Nature **484**, 489 (2012).

[9] D. M. Zajac, T. M. Hazard, X. Mi, E. Nielsen, and J. R. Petta, Phys. Rev. Applied **6**, 054013 (2016).

[10] U. Mukhopadhyay, J. P. Dehollain, C. Reichl, W. Wegscheider, and L. M. Vandersypen, arXiv preprint arXiv:1802.05446 (2018).

[11] R. Barends, J. Kelly, A. Megrant, A. Veitia, D. Sank, E. Jeffrey, T. C. White, J. Mutus, A. G. Fowler, B. Campbell, Y. Chen, Z. Chen, B. Chiaro, A. Dunsworth, C. Neill, P. O'Malley, P. Roushan, A. Vainsencher, J. Wenner, A. N. Korotkov, A. N. Cleland, and J. M. Martinis, Nature **508**, 500 (2014).

[12] M. W. Johnson, M. H. Amin, S. Gildert, T. Lanting, F. Hamze, N. Dickson, R. Harris, A. J. Berkley, J. Johansson, P. Bunyk, E. M. Chapple, C. Enderud, J. P. Hilton, K. Karimi, E. Ladizinsky, N. Ladizinsky, T. Oh, I. Perminov, C. Rich, M. C. Thom, E. Tolkacheva, C. J. Truncik, S. Uchaikin, J. Wang, B. Wilson, and G. Rose, Nature **473**, 194 (2011).

[13] T. Hime, P. Reichardt, B. Plourde, T. Robertson, C.-E. Wu, A. Ustinov, and J. Clarke, Science **314**, 1427 (2006).

[14] A. Niskanen, K. Harrabi, F. Yoshihara, Y. Nakamura, S. Lloyd, and J. Tsai, Science **316**, 723 (2007).

[15] S. H. W. van der Ploeg, A. Izmalkov, A. M. van den Brink, U. Hübner, M. Grajcar, E. Il'ichev, H.-G. Meyer, and A. M. Zagoskin, Phys. Rev. Lett. **98**, 057004 (2007).

[16] R. Harris, A. J. Berkley, M. W. Johnson, P. Bunyk, S. Govorkov, M. C. Thom, S. Uchaikin, A. B. Wilson, J. Chung, E. Holtham, J. D. Biamonte, A. Y. Smirnov, M. H. S. Amin, and A. Maassen van den Brink, Phys. Rev. Lett. **98**, 177001 (2007).

[17] M. S. Allman, F. Altomare, J. D. Whittaker, K. Cicak, D. Li, A. Sirois, J. Strong, J. D. Teufel, and R. W. Simmonds, Phys. Rev. Lett. **104**, 177004 (2010).

[18] S. J. Srinivasan, A. J. Hoffman, J. M. Gambetta, and A. A. Houck, Phys. Rev. Lett. **106**, 083601 (2011).

[19] Y. Chen, C. Neill, P. Roushan, N. Leung, M. Fang, R. Barends, J. Kelly, B. Campbell, Z. Chen, B. Chiaro, A. Dunsworth, E. Jeffrey, A. Megrant, J. Y. Mutus, P. J. J. O'Malley, C. M. Quintana, D. Sank, A. Vainsencher, J. Wenner, T. C. White, M. R. Geller, A. N. Cleland, and J. M. Martinis, Phys. Rev. Lett. **113**, 220502 (2014).

[20] A. Baust, E. Hoffmann, M. Haeberlein, M. J. Schwarz, P. Eder, J. Goetz, F. Wulschner, E. Xie, L. Zhong, F. Quijandría, B. Peropadre, D. Zueco, J.-J. García Ripoll, E. Solano, K. Fedorov, E. P. Menzel, F. Deppe, A. Marx, and R. Gross, Phys. Rev. B **91**, 014515 (2015).

[21] S. J. Weber, G. O. Samach, D. Hover, S. Gustavsson, D. K. Kim, A. Melville, D. Rosenberg, A. P. Sears, F. Yan, J. L. Yoder, W. D. Oliver, and A. J. Kerman, Phys. Rev. Applied **8**, 014004 (2017).

[22] M. Kjaergaard, *in preparation* (2018).

[23] A. Blais, J. Gambetta, A. Wallraff, D. I. Schuster, S. M. Girvin, M. H. Devoret, and R. J. Schoelkopf, Phys. Rev. A **75**, 032329 (2007).

[24] S. Bravyi, D. P. DiVincenzo, and D. Loss, Annals of physics **326**, 2793 (2011).

[25] D. Rosenberg, D. Kim, R. Das, D. Yost, S. Gustavsson, D. Hover, P. Krantz, A. Melville, L. Racz, G. O. Samach, S. J. Weber, F. Yan, J. Yoder, A. J. Kerman, and W. D. Oliver, npj Quantum Information **3**, 42 (2017).

[26] J. Koch, T. M. Yu, J. Gambetta, A. A. Houck, D. I. Schuster, J. Majer, A. Blais, M. H. Devoret, S. M. Girvin, and R. J. Schoelkopf, Phys. Rev. A **76**, 042319 (2007).

[27] M. D. Hutchings, J. B. Hertzberg, Y. Liu, N. T. Bronn, G. A. Keefe, M. Brink, J. M. Chow, and B. L. T. Plourde, Phys. Rev. Applied **8**, 044003 (2017).

[28] M. Steffen, S. Kumar, D. P. DiVincenzo, J. R. Rozen, G. A. Keefe, M. B. Rothwell, and M. B. Ketchen, Phys. Rev. Lett. **105**, 100502 (2010).

[29] F. Yan, S. Gustavsson, A. Kamal, J. Birenbaum, A. P. Sears, D. Hover, T. J. Gudmundsen, D. Rosenberg, G. Samach, S. Weber, J. L. Yoder, T. P. Orlando, J. Clarke, A. J. Kerman, and W. D. Oliver, Nature Communications **7**, 12964 (2016).

[30] Y.-H. Lin, L. B. Nguyen, N. Grabon, J. S. Miguel, N. Pankratova, and V. E. Manucharyan, arXiv preprint arXiv:1705.07873 (2017).

[31] J. M. Chow, A. D. Córcoles, J. M. Gambetta, C. Rigetti, B. R. Johnson, J. A. Smolin, J. R. Rozen, G. A. Keefe, M. B. Rothwell, M. B. Ketchen, and M. Steffen, Phys. Rev. Lett. **107**, 080502 (2011).

[32] D. C. McKay, S. Filipp, A. Mezzacapo, E. Magesan, J. M. Chow, and J. M. Gambetta, Phys. Rev. Applied **6**, 064007 (2016).

[33] M. H. Devoret, Les Houches, Session LXIII **7** (1995).

[34] S. M. Girvin, Proceedings of the 2011 Les Houches Summer School (2011).

[35] See supplementary materials.

[36] J. Majer, J. M. Chow, J. M. Gambetta, J. Koch, B. R. Johnson, J. A. Schreier, L. Frunzio, D. I. Schuster, A. A. Houck, A. Wallraff, A. Blais, M. H. Devoret, S. M. Girvin, and R. J. Schoelkopf, Nature **449**, 443 (2007).

[37] L. DiCarlo, J. M. Chow, J. M. Gambetta, L. S. Bishop, B. R. Johnson, D. I. Schuster, J. Majer, A. Blais, L. Frunzio, S. M. Girvin, and R. J. Schoelkopf, Nature **460**, 240 (2009).

[38] R. C. Bialczak, M. Ansmann, M. Hofheinz, E. Lucero, M. Neeley, A. D. Oconnell, D. Sank, H. Wang, J. Wenner, M. Steffen, A. N. Cleland, and J. M. Martinis, Nature Physics **6**, 409 (2010).

[39] P. J. J. O'Malley, J. Kelly, R. Barends, B. Campbell, Y. Chen, Z. Chen, B. Chiaro, A. Dunsworth, A. G. Fowler, I.-C. Hoi, E. Jeffrey, A. Megrant, J. Mutus, C. Neill, C. Quintana, P. Roushan, D. Sank, A. Vainsencher, J. Wenner, T. C. White, A. N. Korotkov, A. N. Cleland, and J. M. Martinis, Phys. Rev. Applied **3**, 044009 (2015).

[40] J. M. Martinis and M. R. Geller, Phys. Rev. A **90**, 022307 (2014).

# A tunable coupling scheme for implementing high-fidelity two-qubit gates

Fei Yan[1],[*] Philip Krantz[1], Youngkyu Sung[1], Morten Kjaergaard[1], Dan Campbell[1],
Joel I.J. Wang[1], Terry P. Orlando[1], Simon Gustavsson[1], and William D. Oliver[1,2,3]

[1]*Research Laboratory of Electronics, Massachusetts Institute of Technology, Cambridge, MA 02139, USA*
[2]*MIT Lincoln Laboratory, 244 Wood Street, Lexington, MA 02420, USA*
[3]*Department of Physics, Massachusetts Institute of Technology, Cambridge, MA 02139, USA*

### CIRCUIT HAMILTONIAN AND QUANTIZATION

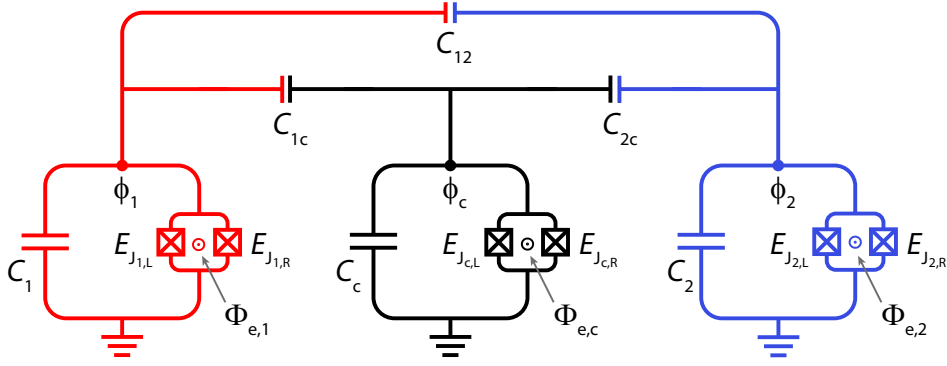

FIG. 1. Circuit diagram of the implemented superconducting circuits, consisting of qubit mode "1" (red), qubit mode "2" (blue) and coupler mode "c" (black). Each mode is a tunable transmon qubit. $E_{J_{\lambda, L(R)}}$ is the Josephson energy of the left(right) junction in mode $\lambda$. $C_\lambda$ is the dominant mode capacitance. $C_{jc}$ ($j = 1, 2$) is the coupling capacitance between qubit $j$ and coupler. $C_{12}$ is the direct coupling capacitance between the two qubits. $\Phi_{e,\lambda}$ is the external magnetic flux threading each loop. $\phi_\lambda$ is the reduced node flux.

The circuit implementing our tunable coupling scheme is shown in Fig. 1. Each transmon qubit may be treated as a weakly anharmonic oscillator consisting of a capacitor $C_\lambda$ and a nonlinear inductance. The inductance is effectively a Josephson junction with a tunable Josephson energy,

$$E_{J_\lambda} = E_{J_{\lambda, \Sigma}} \cos\left(\frac{\pi \Phi_{e,\lambda}}{\Phi_0}\right) \sqrt{1 + d_\lambda^2 \tan^2\left(\frac{\pi \Phi_{e,\lambda}}{\Phi_0}\right)} \,, \tag{1}$$

where $\Phi_0 = h/2e$ is the superconducting flux quantum, $E_{J_{\lambda, \Sigma}} = E_{J_{\lambda, L}} + E_{J_{\lambda, R}}$ is the sum of the Josephson energies and $d_\lambda = \frac{E_{J_{\lambda, L}} - E_{J_{\lambda, R}}}{E_{J_{\lambda, R}} + E_{J_{\lambda, L}}}$ is the junction asymmetry [1]. For simplicity, the self capacitance of Josephson junctions has been incorporated into $C_\lambda$.

We choose node fluxes $\phi_\lambda$ (denoted in Fig. 1) as the generalized coordinates of the system [2, 3]. The system Lagrangian is

$$L = T - U \,, \tag{2}$$

$$T = \frac{1}{2}\left(\frac{\Phi_0}{2\pi}\right)^2 [C_1 \dot{\phi}_1^2 + C_c \dot{\phi}_c^2 + C_2 \dot{\phi}_2^2 + C_{1c}(\dot{\phi}_1 - \dot{\phi}_c)^2 + C_{2c}(\dot{\phi}_2 - \dot{\phi}_c)^2 + C_{12}(\dot{\phi}_1 - \dot{\phi}_2)] \,, \tag{3}$$

$$U = E_{J_1}(1 - \cos\phi_1) + E_{J_c}(1 - \cos\phi_c) + E_{J_2}(1 - \cos\phi_2) \,, \tag{4}$$

where $T$ and $U$ are respectively the kinetic and potential energy. The kinetic energy can be rewritten in a compact form as $T = \frac{1}{2}(\frac{\Phi_0}{2\pi})^2 \dot{\vec{\phi}}^T \mathbf{C} \dot{\vec{\phi}}$, where $\vec{\phi} = [\phi_1, \phi_c, \phi_2]$ and $\mathbf{C}$ is a 3×3 capacitance matrix:

$$\mathbf{C} = \begin{bmatrix} C_1 + C_{1c} + C_{12} & -C_{1c} & -C_{12} \\ -C_{1c} & C_c + C_{1c} + C_{2c} & -C_{2c} \\ -C_{12} & -C_{2c} & C_2 + C_{2c} + C_{12} \end{bmatrix} \tag{5}$$

From the Lagrangian, the generalized momenta $q_\lambda$ – canonical conjugates to the node fluxes – are the node charges

$$q_\lambda = \frac{\partial L}{\partial \dot{\phi}_\lambda} \,, \tag{6}$$

and we have $\vec{q} = (\frac{\Phi_0}{2\pi})^2 \mathbf{C} \dot{\vec{\phi}}$. The classical Hamiltonian can be expressed as

$$H = \sum_\lambda q_\lambda \dot{\phi}_\lambda - L = \frac{1}{2}\vec{q}^T[\mathbf{C}^{-1}]\vec{q} + U \,, \tag{7}$$

where $\mathbf{C}^{-1}$ is the inverse capacitance matrix.

$$\mathbf{C}^{-1} = \frac{1}{||\mathbf{C}||}\begin{bmatrix} A_{11} & A_{12} & A_{13} \\ A_{21} & A_{22} & A_{23} \\ A_{31} & A_{32} & A_{33} \end{bmatrix} \approx \begin{bmatrix} 1/C_1 & C_{1c}/(C_1 C_c) & [C_{12} + (C_{1c}C_{2c})/C_c]/(C_1 C_2) \\ C_{1c}/(C_1 C_c) & 1/C_c & C_{2c}/(C_c C_2) \\ [C_{12} + (C_{1c}C_{2c})/C_c]/(C_1 C_2) & C_{2c}/(C_c C_2) & 1/C_2 \end{bmatrix} \tag{8}$$

$$||\mathbf{C}|| = C_1 C_c C_2 + C_1 C_c C_{2c} + C_c C_2 C_{12} + C_1 C_2 C_{1c}$$
$$+ (C_1 + C_2 + C_c)(C_{1c}C_{2c} + C_{2c}C_{12} + C_{12}C_{1c}) \approx C_1 C_c C_2 \tag{9}$$

$$A_{11} = C_2 C_c + C_2(C_{1c} + C_{2c}) + C_c(C_{2c} + C_{12}) + C_{1c}C_{2c} + C_{2c}C_{12} + C_{12}C_{1c} \approx C_2 C_c \tag{10}$$

$$A_{22} = C_1 C_2 + C_1(C_{12} + C_{2c}) + C_2(C_{12} + C_{1c}) + C_{1c}C_{2c} + C_{2c}C_{12} + C_{12}C_{1c} \approx C_1 C_2 \tag{11}$$

$$A_{33} = C_1 C_c + C_1(C_{1c} + C_{2c}) + C_c(C_{12} + C_{1c}) + C_{1c}C_{2c} + C_{2c}C_{12} + C_{12}C_{1c} \approx C_1 C_c \tag{12}$$

$$A_{12} = A_{21} = C_2 C_{1c} + (C_{12}C_{1c} + C_{1c}C_{2c} + C_{2c}C_{12}) \approx C_2 C_{1c} \tag{13}$$

$$A_{23} = A_{32} = C_1 C_{2c} + (C_{12}C_{1c} + C_{1c}C_{2c} + C_{2c}C_{12}) \approx C_1 C_{2c} \tag{14}$$

$$A_{31} = A_{13} = C_c C_{12} + C_{1c}C_{2c} + (C_{12}C_{1c} + C_{2c}C_{12}) \approx C_c C_{12} + C_{1c}C_{2c} \tag{15}$$

In Eqs. (8-15), we assume that the qubit-coupler coupling capacitances are smaller than the mode any mode capacitance but bigger than the qubit-qubit coupling capacitance. That is, $C_{12} \ll C_{jc} \ll C_\lambda$. However, the magnitude of the factor $\eta = \frac{C_{1c}C_{2c}}{C_{12}C_c}$ is unspecified and can be on the order of unity, so the two terms on the r.h.s. of Eq. (15) can be comparable.

Using canonical quantization, we obtain the quantum-mechanical Hamiltonian,

$$\hat{H} = 4E_{C_1}(\hat{n}_1)^2 - E_{J_1}\cos(\hat{\phi}_1) + 4E_{C_c}(\hat{n}_c)^2 - E_{J_c}\cos(\hat{\phi}_c) + 4E_{C_2}(\hat{n}_2)^2 - E_{J_2}\cos(\hat{\phi}_2)$$
$$+ 8\frac{C_{1c}}{\sqrt{C_1 C_c}}\sqrt{E_{C_1}E_{C_c}}(\hat{n}_1 \hat{n}_c) + 8\frac{C_{2c}}{\sqrt{C_2 C_c}}\sqrt{E_{C_2}E_{C_c}}(\hat{n}_2 \hat{n}_c) + 8(1+\eta)\frac{C_{12}}{\sqrt{C_1 C_2}}\sqrt{E_{C_1}E_{C_2}}(\hat{n}_1 \hat{n}_2) \tag{16}$$

where the operator $\hat{n}_\lambda = -\mathrm{i}\partial/\partial\phi_\lambda$ is the Cooper-pair number operator and $E_{C_\lambda} = \frac{e^2}{2C_\lambda}$ is the charging energy of the corresponding mode.

In the transmon regime, $E_{J_\lambda}/E_{C_\lambda} \gg 1$, the system can be described in the form of coupled Duffing oscillators ($\hbar = 1$):

$$\hat{H} = \hat{H}_1 + \hat{H}_c + \hat{H}_2 + \hat{H}_{1c} + \hat{H}_{2c} + \hat{H}_{12} \tag{17}$$

$$\hat{H}_\lambda = \omega_\lambda \hat{b}_\lambda^\dagger \hat{b}_\lambda + \frac{\alpha_\lambda}{2}\hat{b}_\lambda^\dagger \hat{b}_\lambda^\dagger \hat{b}_\lambda \hat{b}_\lambda \,, \quad \lambda \in \{1, c, 2\} \tag{18}$$

$$\hat{H}_{jc} = g_j(\hat{b}_j^\dagger \hat{b}_c + \hat{b}_j \hat{b}_c^\dagger - \hat{b}_j^\dagger \hat{b}_c^\dagger - \hat{b}_j \hat{b}_c) \,, \quad j = 1, 2 \tag{19}$$

$$\hat{H}_{12} = g_{12}(\hat{b}_1^\dagger \hat{b}_2 + \hat{b}_1 \hat{b}_2^\dagger) \,, \tag{20}$$

where $\hat{b}_\lambda(\hat{b}_\lambda^\dagger)$ denotes the annihilation (creation) operator for the corresponding mode and

$$\omega_\lambda = \sqrt{8E_{J_\lambda}E_{C_\lambda}} - E_{C_\lambda} \,, \tag{21}$$

$$\alpha_\lambda = -E_{C_\lambda} \,, \tag{22}$$

$$g_j = \frac{1}{2}\frac{C_{jc}}{\sqrt{C_j C_c}}\sqrt{\omega_j \omega_c} \,, \tag{23}$$

$$g_{12} = \frac{1}{2}(1+\eta)\frac{C_{12}}{\sqrt{C_1 C_2}}\sqrt{\omega_1 \omega_2} \,. \tag{24}$$

$\omega_\lambda = \omega_{01,\lambda}$ is the oscillator frequency; $\alpha_\lambda = \omega_{12,\lambda} - \omega_{01,\lambda}$ is the oscillator anharmonicity; $g_j$ and $g_{12}$ are respectively the qubit-coupler and qubit-qubit coupling strength. Note that, in Eq. (19), we keep not only the usual Jaynes-Cummings interaction term $(\hat{b}_j^\dagger \hat{b}_c + \hat{b}_j \hat{b}_c^\dagger)$, but also the counter-rotating term $(\hat{b}_j^\dagger \hat{b}_c^\dagger + \hat{b}_j \hat{b}_c)$, because, as we shall discuss below, the contribution from the double-excitation (de-excitation) interaction can also be significant when the coupler frequency becomes much greater than the qubit frequency.

## SCHRIEFFER-WOLFF TRANSFORMATION

To decouple the coupler from the system, we diagonalize the Hamiltonian by using the Schrieffer-Wolff transformation

$$\hat{U} = \exp\left(\sum_{j=1,2}\left[\frac{g_j}{\Delta_j}(\hat{b}_j^\dagger \hat{b}_c - \hat{b}_j \hat{b}_c^\dagger) - \frac{g_j}{\Sigma_j}(\hat{b}_j^\dagger \hat{b}_c^\dagger - \hat{b}_j \hat{b}_c)\right]\right), \tag{25}$$

where $\Delta_j = \omega_j - \omega_c$ and $\Sigma_j = \omega_j + \omega_c$. Compared to the transformation operator used in Ref. [1], we add the second term part in Eq. (25), which accounts for the diagonalization to the counter-rotating terms. In addition, we assume weak anharmonicity, i.e., $\alpha_\lambda \ll \Delta_j$, and use a uniform value $\Delta_j$ for estimating the frequency detuning. Expanding $\hat{U}\hat{H}\hat{U}^\dagger$ in the order of $g_1^{k_1} g_2^{k_2} g_{12}^{k_3}$ and keeping terms up to second order, i.e., $k_1 + k_2 + 2k_3 \leq 2$ ($g_{12}$ is a considered second-order small quantity), we obtain the effective qubit-qubit Hamiltonian

$$\begin{aligned}\tilde{\hat{H}} &= \hat{U}\hat{H}\hat{U}^\dagger \\ &= \tilde{\omega}_1 \hat{b}_1^\dagger \hat{b}_1 + \frac{\tilde{\alpha}_1}{2}\hat{b}_1^\dagger \hat{b}_1^\dagger \hat{b}_1 \hat{b}_1 + \tilde{\omega}_c \hat{b}_c^\dagger \hat{b}_c + \frac{\tilde{\alpha}_c}{2}\hat{b}_c^\dagger \hat{b}_c^\dagger \hat{b}_c \hat{b}_c + \tilde{\omega}_2 \hat{b}_2^\dagger \hat{b}_2 + \frac{\tilde{\alpha}_2}{2}\hat{b}_2^\dagger \hat{b}_2^\dagger \hat{b}_2 \hat{b}_2 + \tilde{g}(\hat{b}_1^\dagger \hat{b}_2 + \hat{b}_1 \hat{b}_2^\dagger),\end{aligned} \tag{26}$$

where

$$\tilde{\omega}_1 \approx \omega_1 + g_1^2\left(\frac{1}{\Delta_1} - \frac{1}{\Sigma_1}\right), \tag{27}$$

$$\tilde{\alpha}_1 \approx \alpha_1, \tag{28}$$

$$\tilde{\omega}_c \approx \omega_c - g_1^2\left(\frac{1}{\Delta_1} + \frac{1}{\Sigma_1}\right) - g_2^2\left(\frac{1}{\Delta_2} + \frac{1}{\Sigma_2}\right), \tag{29}$$

$$\tilde{\alpha}_c \approx \alpha_c, \tag{30}$$

$$\tilde{\omega}_2 \approx \omega_2 + g_2^2\left(\frac{1}{\Delta_2} - \frac{1}{\Sigma_2}\right), \tag{31}$$

$$\tilde{\alpha}_2 \approx \alpha_2, \tag{32}$$

$$\tilde{g} \approx \frac{g_1 g_2}{2}\left(\frac{1}{\Delta_1} + \frac{1}{\Delta_2} - \frac{1}{\Sigma_1} - \frac{1}{\Sigma_2}\right) + g_{12}. \tag{33}$$

In Eq. 26, we have assumed the coupler is at ground state ($\hat{b}_c^\dagger \hat{b}_c = 0$) and $\alpha_\lambda$ is also a small quantity. In the strong dispersive regime ($\omega_c \gg \omega_j$), $|\Delta_j| \approx |\Sigma_j|$, so the counter-rotating term does contribute significantly. The computational states $|100\rangle$ and $|001\rangle$ exchange their energy virtually through the Jaynes-Cummings interaction $(\hat{b}_j^\dagger \hat{b}_c + \hat{b}_j \hat{b}_c^\dagger)$ via the non-computational state $|010\rangle$, and also through the counter-rotating term $(\hat{b}_j^\dagger \hat{b}_c^\dagger + \hat{b}_j \hat{b}_c)$ via a higher non-computational state $|111\rangle$.

Finally, substituting Eq. (23-24) into Eq. (33), we have

$$\tilde{g} = \frac{1}{2}\left[\frac{\omega_c}{4}\left(\frac{1}{\Delta_1} + \frac{1}{\Delta_2} - \frac{1}{\Sigma_1} - \frac{1}{\Sigma_2}\right)\eta + \eta + 1\right]\frac{C_{12}}{\sqrt{C_1 C_2}}\sqrt{\omega_1 \omega_2}, \tag{34}$$

which recovers Eq. (5) in the main text. Assuming $\omega_1 = \omega_2 = \omega$,

$$\tilde{g} = \frac{1}{2}\left[\frac{\omega^2}{\omega^2 - \omega_c^2}\eta + 1\right]\frac{C_{12}}{\sqrt{C_1 C_2}}\sqrt{\omega_1 \omega_2}. \tag{35}$$

The first term in the bracket vanishes when $\omega_c$ goes to infinity. Therefore, given arbitrarily small $C_{12}$, there is a guaranteed solution for $\omega_c$ such that $\widetilde{g} = 0$.

---

[1] J. Koch, T. M. Yu, J. Gambetta, A. A. Houck, D. I. Schuster, J. Majer, A. Blais, M. H. Devoret, S. M. Girvin, and R. J. Schoelkopf, Phys. Rev. A **76**, 042319 (2007).
[2] M. H. Devoret, Les Houches, Session LXIII **7** (1995).
[3] S. M. Girvin, Proceedings of the 2011 Les Houches Summer School (2011).